\documentclass[conference]{IEEEtran}
\IEEEoverridecommandlockouts
\usepackage{cite}
\usepackage{amsmath,amssymb,amsfonts}
\usepackage{graphicx}
\usepackage{textcomp}
\usepackage{subfigure}
\usepackage{algorithm}
\usepackage[noend]{algpseudocode}
\usepackage[normalem]{ulem}
\usepackage{fancyhdr}
\usepackage{dsfont}
\usepackage{comment}
\usepackage{enumitem}
\usepackage{url}
\usepackage[lmargin=0.7in, rmargin=0.7in, tmargin=0.7in, bmargin=1in]{geometry}
\usepackage{multirow}
\usepackage{rotating}
\usepackage[table,xcdraw]{xcolor}
\usepackage[normalem]{ulem}
\usepackage{tabularray}
\usepackage[framemethod=TikZ]{mdframed}
\usepackage[english]{babel}
\usepackage{amsthm}

\newtheorem{remark}{Remark}

\newmdtheoremenv[
leftmargin=3pt,
rightmargin=3pt,
backgroundcolor=gray!40,
roundcorner=4pt,
theoremtitlefont=\bfseries,
]{obsv}{Observation}

\def\BibTeX{{\rm B\kern-.05em{\sc i\kern-.025em b}\kern-.08em
    T\kern-.1667em\lower.7ex\hbox{E}\kern-.125emX}}
\begin{document}

\title{
The Robustness of Spiking Neural Networks in Federated Learning with Compression Against Non-omniscient Byzantine Attacks 
}


 \author{\IEEEauthorblockN{
 Manh V. Nguyen\IEEEauthorrefmark{1},
 Liang Zhao\IEEEauthorrefmark{1},
 Bobin Deng\IEEEauthorrefmark{1},
 Shaoen Wu\IEEEauthorrefmark{1}}
 \IEEEauthorblockA{\IEEEauthorrefmark{1}College of Computing and Software Engineering,
 Kennesaw State University, Georgia, USA
 \\mnguy126@students.kennesaw.edu, \{lzhao10, bdeng2, swu10\}@kennesaw.edu}
 }

\maketitle

\begin{abstract}

Spiking Neural Networks (SNNs), which offer exceptional energy efficiency for inference, and Federated Learning (FL), which offers privacy-preserving distributed training, is a rising area of interest that highly beneficial towards Internet of Things (IoT) devices. 
Despite this, research that tackles Byzantine attacks and bandwidth limitation in FL-SNNs, both poses significant threats on model convergence and training times, still remains largely unexplored. 
Going beyond proposing a solution for both of these problems, in this work we highlight the dual benefits of FL-SNNs -- against non-omniscient Byzantine adversaries (ones that restrict attackers access to local clients' datasets), and greater communication efficiency -- over FL-ANNs. Specifically, we discovered that a simple integration of Top-$\kappa$ sparsification into the FL apparatus can help leverage the advantages of the SNN models in both greatly reducing bandwidth usage and significantly boosting the robustness of FL training against non-omniscient Byzantine adversaries. Most notably, we saw a massive improvement of $\sim$40\% accuracy gain in FL-SNNs training under the lethal MinMax attack.



\end{abstract}

\begin{IEEEkeywords}
FL, SNNs, Byzantine, Top-k sparsification
\end{IEEEkeywords}

\section{Introduction}

The quick evolution of AI led to the data center increasing more than 100X the number of accelerators and, therefore, the power overhead that meets our continuous AI growth becomes a mission-critical issue~\cite{beth2024aipower}. Georgia Power predicted that the summer demand in 2031 will be 28 times more than the projected in its 2022 cases, and the energy-sucking data centers are one of the main factors that are driving the demand~\cite{kann2024hear}. Therefore, Georgia Power plans to burn more fossil fuels to fill this energy gap. From the AI model side, Spiking Neural Networks (SNNs) have emerged as a promising solution for next-generation AI due to their potential for high energy efficiency, which provides an excellent opportunity to maintain AI capability growth and environment sustainability. Specifically, SNNs mimic biological neurons in the human brain to provide energy-efficient machine learning systems with attributes including event-driven (sparsity), real-time local processing, and temporal data storage. These attributes offer many new attractive user cases for edge systems (e.g., autonomous driving systems), such as energy-efficient smart IoT devices and event cameras). SNN in-situ learning, i.e., directly on-device model training, is a hotly pursued topic associated with smart edge systems for better privacy preservation, communication bandwidth, and latency. However, the limited computational and storage resources on IoT devices also constrained the feasibility of SNN in-situ training on edge. To aid this effort, Federated Learning (FL) emerges as a collaborative learning framework that provides distributed training on the edge while protecting the data privacy of edge devices. Therefore,  FL delivers a foundation for developing SNN in-situ training on edge. 


Byzantine adversaries pose a significant threat to FL systems in the form of undetected malicious clients. 
Specifically, they seek to manipulate the submitted local model updates to introduce harm to the global model. 
There are two modes of model poisoning attacks: 1) \textit{non-omniscient}, where the adversaries do not have access to the training data, and 2) \textit{omniscient} otherwise \cite{baruch2019little}. 
In this work, we focus on the non-omniscient types of attack, where at most, the Byzantine clients assume knowledge of all the parameter updates from the benign clients. 
With this awareness, the more sophisticated adversaries aim to disguise the malicious updates from statistically robust aggregation methods \cite{yin2018byzantine}. 
The state-of-the-art of these kinds is the MinMax attack, which optimizes the malicious updates based on distance between the two benign updates furthest from each others \cite{shejwalkar2021manipulating}. 

Another concern in FL training is bandwidth limitation, as most IoT devices communicate via unstable wireless networks with much smaller and more heterogeneity in networking capacity. 
This can create server-side bottlenecks, or straggling at the lower bandwidth devices that result in prolonged FL training time. 
On the other hand, back-propagation through time (BPTT) with surrogate gradients \cite{neftci2019surrogate} has become a promising in situ training methodology for SNNs, such a different training technique produces statistical differences between the parameters of SNN models and those of equivalent ANN models. 
As non-omniscient Byzantine disguises their malicious gradients behind the benign gradients' statistics, these differences can potentially be used to enhance the robustness of FL-SNNs. Therefore, this work experiments with Top-k sparsification, as a gradient compression technique to reduce communication overhead, but also as a mean to capitalize on this robustness potential of SNNs.

\textbf{Contribution.} 
In this work, beyond merely providing a Byzantine defense or a bandwidth optimization methodology for FL, we also explore the advantages of SNNs over ANNs in these specific contexts. Our key contribution is highlighted in three-fold: 
\textit{i)} To the best of our knowledge, this is the first work that explores the robustness of FL-SNN compared to FL-ANN under Byzantine attacks and bandwidth limitation. We conduct the FL training of both models against four relevant Byzantine adversaries and discover that FL-SNN is more robust on FL-ANN under these attacks, that is, except for the MinMax adversary; 
\textit{ii)} We put-forth the prospects of communication efficiency of FL-SNN with Top-$\kappa$ sparsification as the model compression methodology; 
\textit{iii)} We conduct the evaluation of FL training with this compression paradigm for SNN and ANN models and observe a massive increase (roughly 40\% accuracy improvement, far surpassing FL-ANNs) in robustness with FL-SNNs against the MinMax Byzantine attack. 
The simple Top-$\kappa$ sparsification technique with SNN model can be used to tackle both non-omniscient Byzantine attacks and bandwidth limitation in FL training.

\section{Related Work} \label{sec:relate}

Several work related to SNNs concerning its utility in FL has been put-forth in recent years. Namely, Venkatesha et al. \cite{venkatesha_federated_2021}, demonstrates the better accuracy and energy efficiency of SNNs over ANNs in large-scale FL; The work Aouedi et al. \cite{aouedi_hfedsnn_2023}, Yu et al. \cite{yu2024heterogeneous} concerns assessing different FL-SNNs training architecture; And Xie et al. \cite{xie_efficient_2022} introduce a new temporal encoding mechanism for FL-SNNs which also demonstrated improved accuracy and energy efficiency on traffic sign recognition data. 
With the overhead of communication being a major issue in FL \cite{shah2021model, haddadpour2021federated}, several papers have addressed it in FL-SNNs training. Namely, Chaki et al. \cite{chaki_communication_2023} explore the accuracy-bandwidth trade-offs; while Liu et al. \cite{liu_federal_2022} and Xie et al. \cite{xie2024federated} demonstrate the advantage of different knowledge-distillation architecture for SNNs in FL training. 
While several work has discussed the vulnerability of FL-SNNs against backdoor attacks -- 
i.e.
Fu et al. \cite{fu_spikewhisper_2024}, and Walter et al. \cite{walter2024mitigating}, which examines the threat of backdoor attacks by exploiting vulnerabilities in neuromorphic data and its defense -- the landscape of SNN models' potential against Byzantine attacks in Fl has yet to be investigated. 
Our work is the first to incorporate FL-SNNs training under both adversarial and communication limited environment.

\section{Background} \label{sec:background}

\subsection{Spiking Neural Networks (SNNs)}

\begin{figure}
    \centering
    \includegraphics[scale=0.65]{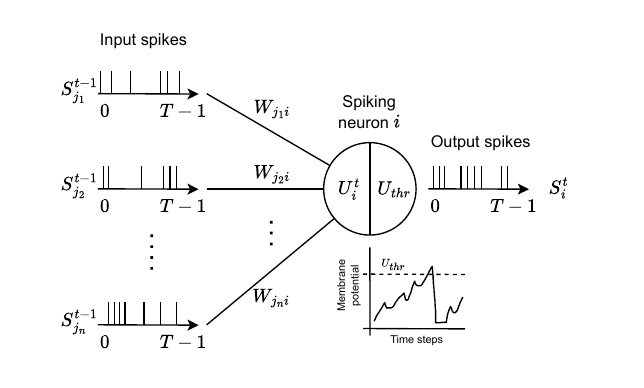}
    \caption{The Leaky Integrate \& Fire (LIF) Spiking Neuron.}
    \label{fig:snn_diagram}
    \vspace{-10pt}
\end{figure}

A spiking neuron simulate the leaky integrate \& fire (LIF) model of the biological neuron, which perceives input as incoming spikes over a predefined time interval \cite{eshraghian2023training}. The neuron accumulates membrane potential while absorbing the input spikes and scales by the synaptic weights. Once the membrane potential reaches a certain threshold, the neuron fires an output spike, releases and reset the membrane energy (Fig. \ref{fig:snn_diagram}). The following equation describes the LIF mechanism of a spiking neuron $i$:
\begin{multline}
U_i^t = \sum_{j \in N} W_{ji} S_{j}^{t-1} + \beta U_i^{t-1} - S_{i}^{t-1} U_{thr},
\\
\text{where } \beta < 1 \text{ and } S_i^{t} = \begin{cases} 
1 & \text{if } U_i^{t} > U_{thr}, \\
0 & \text{otherwise}. \end{cases}
\end{multline}
in which, $S_i^t$ is the binary output of neuron $i$, and $U_i^t$ is its membrane potential at timestep $t$. 
While $\beta$ represents the leak factor by which the membrane potential is reduced at each time step and $U_{thr}$ represents the membrane threshold. $N$ denotes the set of input neurons that is connected to $i$, and $W_{ji}$ denotes the synaptic weight of the $j\rightarrow i$ connection. In this work, SNNs is trained via back-propagation through time (BPTT) \cite{lee2016training}. To address the problem of non-differentiable step function with threshold, Neftci et al. propose a surrogate function as an approximation for the spiking function's gradient \cite{neftci2019surrogate, bohte2011error}.

\subsection{Federated Learning (FL)}

\begin{figure}
    \centering
    \includegraphics[width=0.8\linewidth]{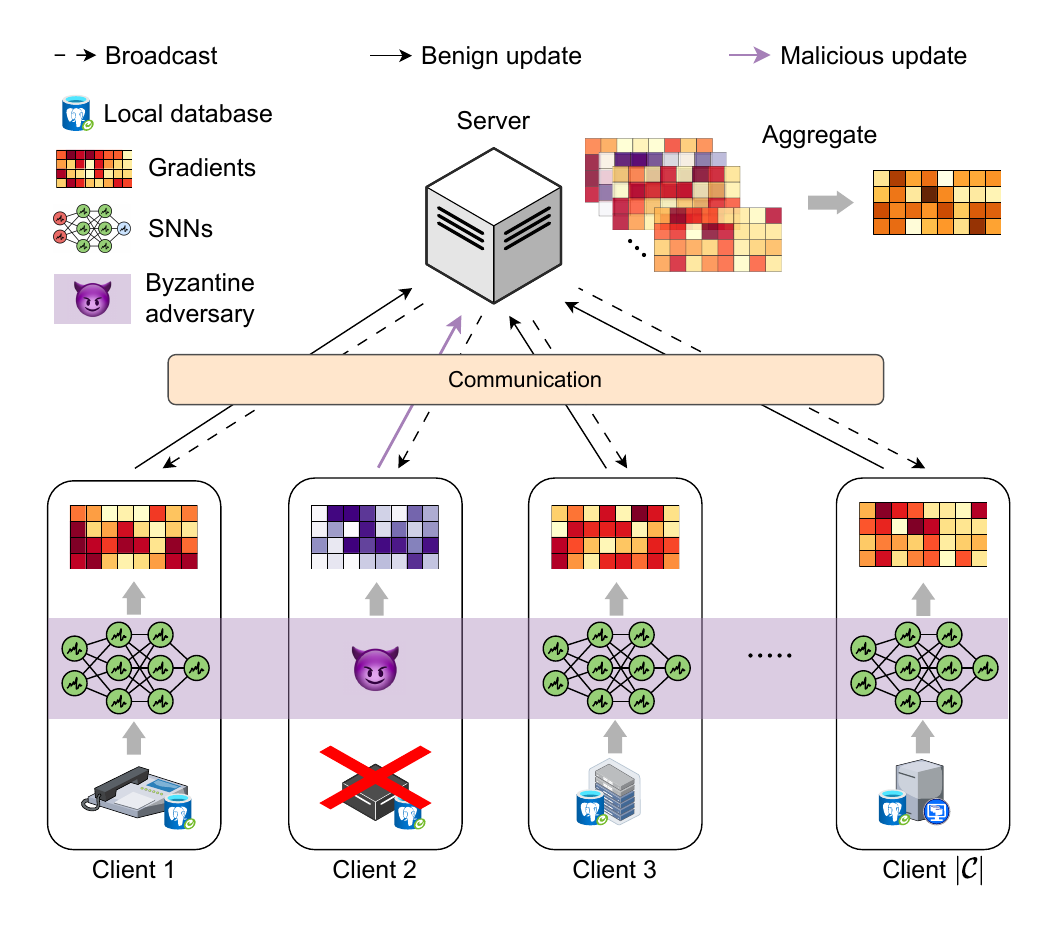} 
    \caption{FL-SNNs Training under Non-omniscient Byzantine Adversary.}
    \label{fig:FL_diagram}
    \vspace{-10pt}
\end{figure}

Federated Learning emerged as a distributed learning solution to centralized the training result, i.e. the model, without compromising the private datasets of the involved parties. This work adopt the \textit{FedAvg} algorithm for simplicity \cite{mcmahan2017communication}, whose aggregation equation is expressed as follows:
\begin{equation}
    \mathbf{W}^{r} \leftarrow \mathbf{W}^{r - 1} + \frac{1}{|\mathcal{C}|} \sum_{c \in \mathcal{C}} \Delta^r_{c} \label{eq:fedavg}
\end{equation}
in which, $r$ denotes the current global training round, $\mathbf{W}_r$ denotes centralized model at $r$, the $\mathcal{C}$ denotes a set of local clients with privy datasets, at $r$, client $c in \mathcal{C}$ produce local model $\mathbf{W}^r_c$ from previous global model $\mathbf{W}_{r-1}$. We denote $\left \{ \Delta^r_c : c \in \mathcal{C}\right\}$, where $\Delta^r_c \leftarrow \mathbf{W}_c^r - \mathbf{W}^{r-1}$, as the gradients of the clients at the current round. 

\subsection{Non-omniscient Byzantine Adversaries} 

\subsubsection{The Byzantine methodology} In FL, a Byzantine attack scheme occurs when supposedly trusted node(s) in the network becomes malicious and actively participating to damage the systems. We  model the submitted gradients from the FL clients under Byzantine attack as follows:
\begin{equation}
\Delta^r_{c} \leftarrow \begin{cases} 
\nabla^m & \text{if $c$-th client is Byzantine}, \\
\mathbf{W}^r_{c} - \mathbf{W}^{r-1} & \text{if $c$-th client is benign}, \end{cases}
\end{equation}
in which, $\nabla^m$ is a malicious value generated with the purpose of tampering with the final aggregated model. 

\subsubsection{The non-omniscient attacks}

In this section, we further explore SNNs' robustness against Non-omniscient Byzantine adversaries. Namely, the four following types: 
\textit{i) Noise \cite{li2021byzantine}:} A straight-forward attack by sampling random noise (usually Gaussian) and adding it on top of the gradients, which is a simple but easily detectable adversary; 
\textit{ii) ALIE \cite{baruch2019little}:} Aiming to introduce subtle and hard-to-detect noise, the attacker first examine the distribution the gradients across the benign updates, the malicious noise is then generated so as to avoid deviating too much from the distribution; 
\textit{iii) MinMax \cite{shejwalkar2021manipulating}:} Improving upon \textit{ALIE}, this adversary maximize the deviating from the true update by exploiting the maximum distance between the benign gradients; 
\textit{iv) IPM \cite{xie2020fall}:} Submit update that negate the true gradients, but in subtle portion to maintain the descent of the training loss, thus keeps the attack unknown.

\begin{figure}
    \centering
    \includegraphics[scale=.4]{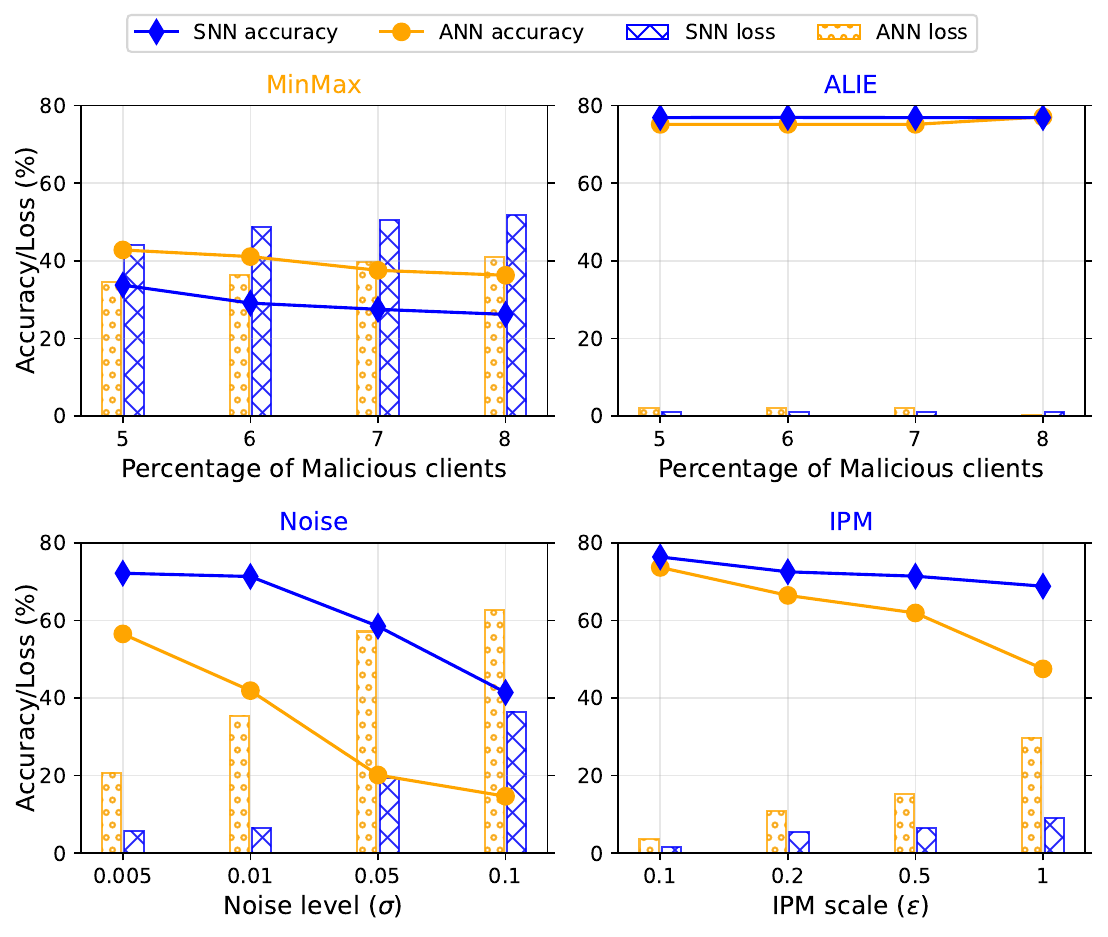}
    \caption{The robustness of FL-SNNs and FL-ANNs against 4 randomized-noise-based Byzantine adversaries. (Note: Loss bars are not to be confused with the training loss, but $\textit{Loss} = \textit{Clean Accuracy} - \textit{Attacked Accuracy}$)}
    \label{fig:prelim_res}
    \vspace{-10pt}
\end{figure}

\section{Pre-Evaluation: FL-SNNs under Non-omniscient Byzantine \& Top-$\kappa$ Sparsification} \label{sec:prelim}

\subsection{Non-omniscient Byzantine Adversaries} \label{subsec:pre_byz_adv}

\subsubsection{Experiments} We benchmark the ANN VGG9 model with its equivalent SNN model, utilizing the dataset CIFAR10 (more details on the models' hyperparameters will be provided in Section \ref{sec:experiment}. 
We perform each FL training among $|\mathcal{C}| = 20$ clients, for $R = 2,000$ rounds, in each round, the local client train for one batch in the dataset. For both models, we establish a baseline clean accuracy (of training without attack), both achieve the final accuracy of roughly $77\%$. 
For each adversarial attack, we compare the robustness of SNN with ANN against the attack in a sub-figure via two metrics: attacked accuracy, and accuracy loss (i.e. $\textit{Clean Accuracy} - \textit{Attacked Accuracy}$). 
For reliability in benchmarking, we vary the intensity of each attack method differently according to its specification.
Specifically, for \textit{Noise}, we generate the random sample using the Gaussian distribution, with mean $\mu = 0$ and varied levels of standard deviation $\sigma \in \{0.005, 0.01, 0.05, 0.1\}$. 
For \textit{IPM}, we vary the value of $\epsilon \in \{0.1, 0.2, 0.5, 1\}$, which is the scale by which the gradients are multiplied by when negated. 
For \textit{ALIE} and \textit{MinMax}, we vary then intensity of attack by the percentage of malicious clients $\frac{|\mathcal{M}_c|}{|\mathcal{C}|} \in \{25\%, 30\%, 35\%, 40\%\}$. 
\subsubsection{Observation} The benchmarking result is displayed in Fig. \ref{fig:prelim_res}, from which we observe that the ALIE attack does not have much effect on the accuracy, this is perhaps because we use FedAvg instead of FedSGD. 
With Noise and IPM attacks, as the intensity of attack increases, we saw that the SNN model is more robust than ANN with the accuracy loss nearly half by that of its ANN counterpart. With regards to the MinMax attack, however, we saw the opposite trend, where the SNN model is less robust than its ANN counterpart, with the accuracy loss of $\sim$$50\%$ compared to $\sim$$40\%$, respectively. From the experiment, we have the following remark:
\begin{remark}
    FL-SNNs generally perform better than FL-ANNs under almost all non-omniscient Byzantine adversaries, except for the MinMax attack.
\end{remark}

\subsection{Federated Learning with Top-$\kappa$ Sparsification} \label{sec:robust}

\begin{figure}
    \centering
    \includegraphics[width=\linewidth]{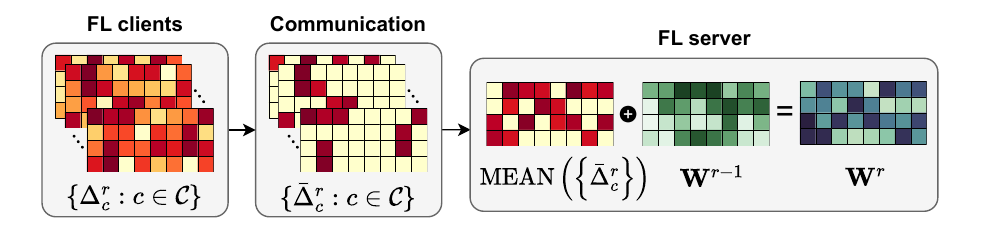}
    \caption{Illustration of FL with Top-$\kappa$ sparsification}
    \label{fig:cmp_proc}
\end{figure}

\begin{figure}
    \centering
    \subfigure[]{\includegraphics[width=.33\linewidth]{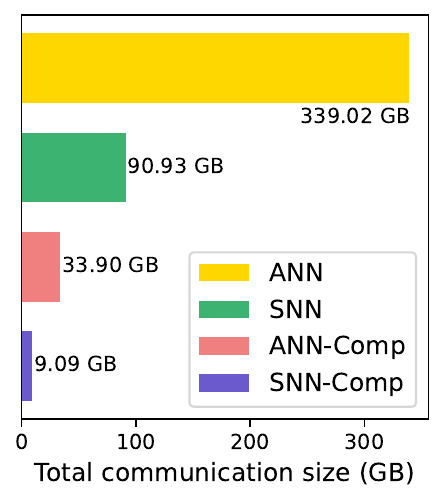}\label{fig:comm_size}}
    \subfigure[]{\includegraphics[width=.5\linewidth]{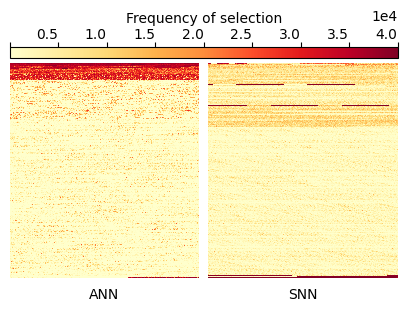}\label{fig:comm_heat}}

    \caption{a) Total bandwidth consumption of FL-SNN and FL-ANN w/ and w/o compression; b) Gradients' frequency of Top-$\kappa$ retention heatmap.}
    \label{fig:comm}
    \vspace{-10pt}
\end{figure}

\subsubsection{Compression mechanism}
Various methods of Top-$k$ based model sparsification has been proposed and analyzed in \cite{shi2019understanding}. 
Which postulate that \textit{gradients with highest magnitudes are the most valuable}. 
Thus, they retain a number of gradients whose magnitudes surpass a certain threshold value. 
We modify this method by retaining a fixed percentage of the highest-magnitude gradients and called this process FL training with Top-$\kappa$ Sparsification (FLTS) \cite{nguyen2024robustness}, in which the compression rate $\kappa \leq 1$ denotes the proportion of retained gradients. 
Fig. \ref{fig:cmp_proc} summarizes this process, in which $\{\Delta^r_c: c \in \mathcal{C}\}$ are the local gradients obtained by the clients after training (or malicious updates if they are Byzantine). In communication, the gradients are compressed into $\{\bar{\Delta}^r_c : c \in \mathcal{C}\}$ via Top-$\kappa$, which are then received and aggregated by the server, whose product, i.e. $\textsc{Mean}\left(\{\bar{\Delta}^r_c\}\right)$, is then added into the current global model $\mathbf{W}^{r-1}$ to make the updated global model $\mathbf{W}^r$, thus comply with Eq. \eqref{eq:fedavg}. 

\subsubsection{Prospects} Fig. \ref{fig:comm} shows us the prospects of communication efficiency of FL-SNNs with Top-$\kappa$ sparsification. 
In which, we set up FL training with and without compression for ANN and SNN models, Fig. \ref{fig:comm_size} displays the total bandwidth usage in each scenario with both models; While in Fig. \ref{fig:comm_heat}, we sample the frequencies of gradients being in the Top-$\kappa$ and retained over the rounds and plot the heat map for this. 
The compression rate (if used) is $\kappa = 0.1$ for both sub-figures. 
From the projection in Fig. \ref{fig:comm_size} we observe that parameters size of SNN, coupled with the compression technique, can bring bandwidth usage down to be $\sim$33 times more efficient than regular FL-ANNs training. From Fig. \ref{fig:comm_heat}, we also observe that the selected gradients are more evened out in the SNN model than its ANN counterpart, this indicates that the global parameters in FL-SNNs are more equally updated than FL-ANNs which shows that Top-$\kappa$ sparsification might works better in FL-SNNs than FL-ANNs.

\section{Performance Evaluation} \label{sec:experiment}

In this section, we evaluate the performance of FL-SNNs integrated with Top-$\kappa$ sparsification. Specifically, Section \ref{subsec:env} details our experiment specification; Section \ref{subsec:eff_comp} demonstrate the robustness of the proposed scheme against the MinMax attack; Section \ref{subsec:reeval} replay the evaluation of Section \ref{subsec:pre_byz_adv}, this time with Top-$\kappa$ integrated.

\subsection{Environment setup} \label{subsec:env}

We evaluate our compression approach using the CIFAR10 for SNN performance evaluations. The training data is evenly distributed among $20$ clients, all clients participated in every global aggregation round.
We employ the VGG9 architecture for both SNN and ANN models. The following table provide the relevant hyper-parameters settings for each model:
\vspace{0pt}
\begin{center}
    \scalebox{0.7}{
    \begin{tabular}{|l|c|c|c|c|c|c|c|}
    \hline
             & Leak $(\beta)$ & Timesteps $(T)$ & Learning & Decay              & Momentum  & Batch  \\ \hline
    ANN      & N/A        & N/A   & 0.0001     & $5\times 10^{-4}$  & $0.9$     & $32$        \\ \hline
    SNN      & 0.99       & 25    & 0.1  & $10^{-4}$          & $0.95$    & $32$        \\ \hline
    \end{tabular}
    }
\end{center}
\vspace{0pt}
\noindent Per global aggregation round, each client trains on $1$ batch of data, a full training consists of a total of $2000$ global rounds.
We utilize the Blade framework \cite{li2024blades} to simulate FL with Byzantine attacks and to parallelize the experiments' executions.
All evaluations were conducted on a centralized server equipped with an 80-core Intel\textsuperscript{\textregistered} Xeon\textsuperscript{\textregistered} CPU E5-2698 v4 @ 2.20GHz processor and 503 GB of memory. The system also featured eight NVIDIA Tesla V100-SXM2 GPUs, each providing 32 GB of memory.

\subsection{Improvement of Robustness via Compression} \label{subsec:eff_comp}

\begin{figure}
    \centering
    \subfigure[]{\includegraphics[scale=.4]{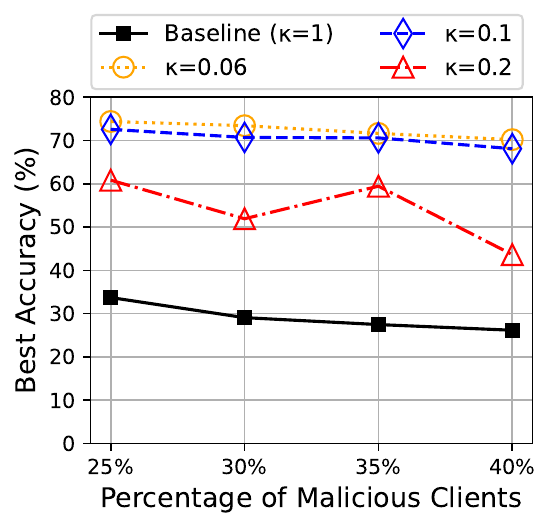}\label{fig:SNN_diff_comp}}
    \subfigure[]{\includegraphics[scale=.4]{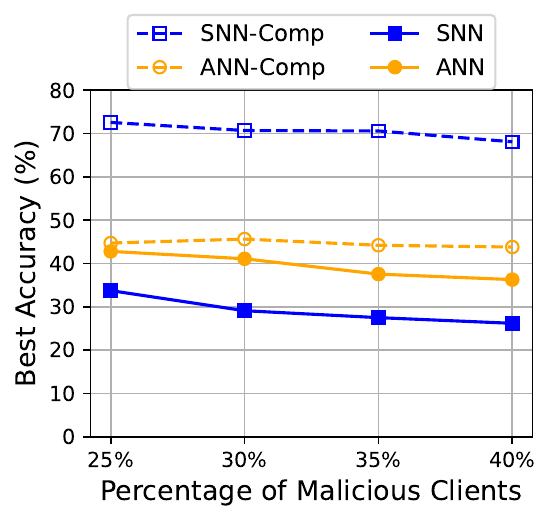}\label{fig:SNN_ANN_comp}}
    \caption{From this figure we can observe that SNN with BRC bring SNN from the worse defense against the MinMax attack to the best one, compared even to ANN, and ANN with BRC}
    \label{fig:SNN_comp_perf}
    \vspace{-10pt}
\end{figure}

In this section, we demonstrate the robustness improvement that Top-$\kappa$ bring against the MinMax attack in Fig. \ref{fig:SNN_comp_perf}. 
Specifically, Fig. \ref{fig:SNN_diff_comp} the robustness of FL-SNNs with different sparsification rate $\kappa \in \{0.06, 0.1, 0.2\}$.
From which, we can see a clear pattern of improvement in FL-SNN with and without the integrated sparsification technique. 
More interestingly, SNN achieves even higher accuracy with a lower sparsification rate, i.e. more compression, and the improvement from $\kappa = 0.2$ to $\kappa = 0.1$ is much vastly than going down lower to $\kappa = 0.06$.
Namely, the parameters in the Top-$0.1$ percentile experience much lower MinMax perturbation than the rest of the parameters, and this effect fades quickly as we increase the percentile beyond $\kappa=0.1$.

On the otherhand, Fig. \ref{fig:SNN_ANN_comp} compares the accuracy between ANN and SNN, with and without compression, for both models, if compression is used, we set $\kappa = 0.1$. 
From which, we can clearly observe the positive impact Top-$\kappa$ sparsification has on the FL robustness of both models against the MinMax attack. 
Specifically, with FL-ANN, we observe a slight improvement in accuracy, i.e. moving from range $37\%-42\%$ to about $47\%$, a roughly $7\%$ increase. 
With FL-SNN, on the other hand, we see a drastic increase in robustness under MinMax by applying the compression method, going from range $28\%-34\%$ to up to range $69\%-73\%$, a massive improvement of roughly $40\%$ increase. Overall, the robustness of FL-SNN significantly improves with the integration of Top-$\kappa$ sparsification. Notably, while robustness increases as the compression rate $\kappa$ decreases, it follows a diminishing return trend, ultimately failing to converge when $\kappa$ becomes excessively low at $0.05$. We have the following remark:
\begin{remark}
    Top-$\kappa$ sparsification brings massive robustness to FL-SNN against the MinMax adversary, however, the same effect is hardly noticeable in FL-ANN.
\end{remark}

\subsection{Re-Evaluation with Non-omniscient Byzantine Adversaries} \label{subsec:reeval}

\begin{figure}
    \centering
    \includegraphics[scale=0.4]{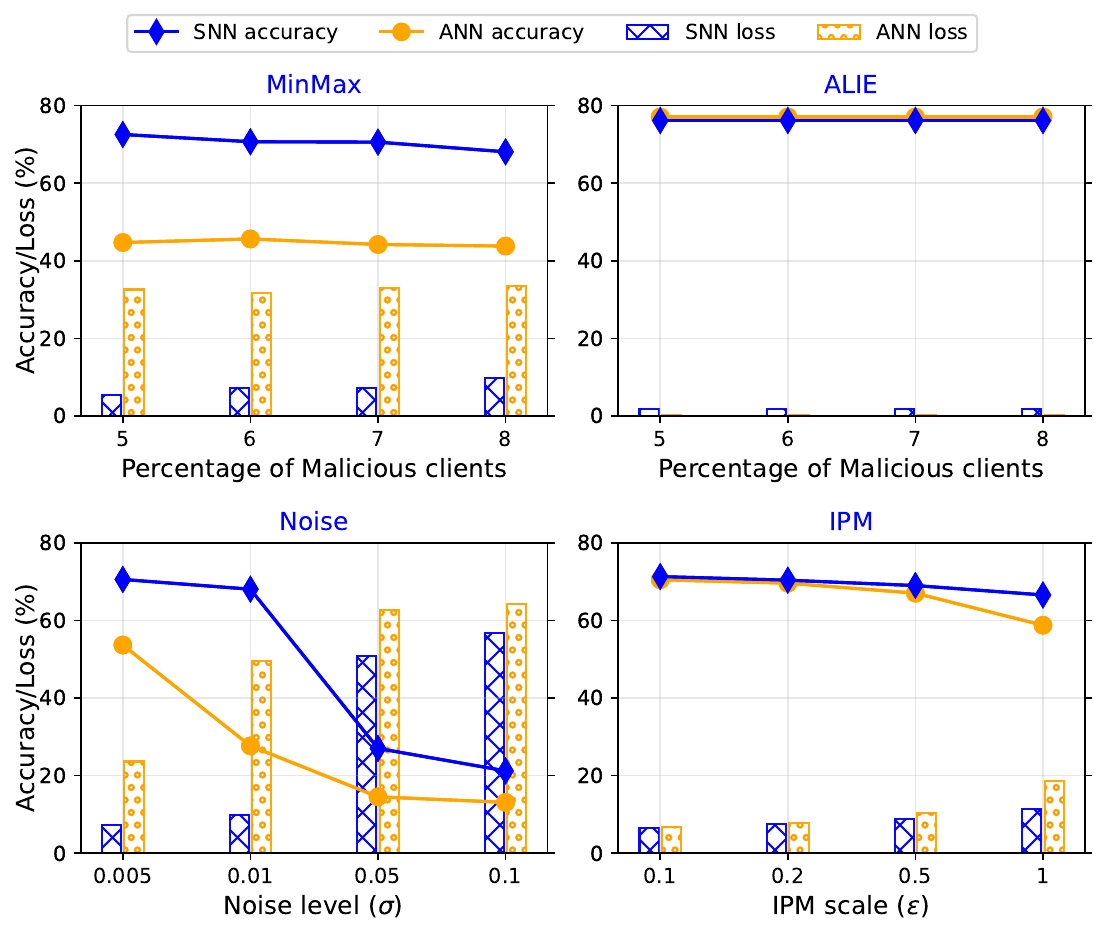}
    \caption{Re-evaluation of FL-SNNs and FL-ANNs with Top-$\kappa$ Sparsification}
    \label{fig:afterwards}
    \vspace{-10pt}
\end{figure}

In the previous section, we find that FL-SNNs under compression increased and is superior to FL-ANNs in robustness against MinMax attack, the questions posed is that how would Top-$\kappa$ compression alter the FL training performance under the rest of the noise-based Byzantine attacks (as presented in Section \ref{sec:prelim}). 
Thus, in this section, we conduct a complete re-evaluation of FL with Top-$\kappa$ sparsification under the same four attacks. 
We present two experiments, in Fig. \ref{fig:afterwards}, we repeat the experiment in pre-evaluation, this time with compression integrated to once again compare the robustness of SNN with ANN, under the same four attacks setting, we set the sparsification rate $\kappa = 0.1$. 
We observe that the accuracy drops quite a bit with the Noise attack when integrated with compression. 
However, the SNN model is still more robust than ANN in this case. 
Furthermore, the Noise attack is easily detectable with statistical examination.
On the other hand, we see that the SNN model becomes much more robust than its ANN counterpart with regards to the dangerous MinMax attack. And overall, we observe that with Top-$\kappa$ compression integrated, SNNs are more robust than ANNs against non-omniscient Byzantine attacks.

\section{Conclusion \& Future Work} \label{sec:conclude}

In summary, going beyond proposing a robust aggregation solution against the MinMax model-poinsoning adversary, our paper explored and demonstrated the robustness of SNN compared to ANN models in the FL context under both Byzantine noise-based adversaries and limited bandwidth conditions.
First, we conduct a pre-evaluation showing that FL-SNNs is more vulnerable to FL-ANNs against the MinMax attack, which is generally not the case for non-omniscient Byzantine adversaries.
Second, we summarize potential Top-$\kappa$ sparsification as the compression method to reduce the communication overhead, especially integrated with FL-SNNs. 
Third, we conduct thorough evaluation of this compression scheme on both models, our experiment showed a massive robustness improvement in FL-SNNs of over $40\%$ of accuracy gain under the MinMax attack, a stark differences when compared to FL-ANNs. 
Finally, we conduct an overall re-evaluation of FL-ANNs and FL-SNNs with the rest of the non-omniscient attacks, the result affirms the superiority of FL-SNNs with Top-$\kappa$ compression in all aspects.
Our future work will aim to further explain the nature SNN models that provided this massive improvement, and potentially leverage it in other defense applications in FL.



\bibliographystyle{IEEEtran}
\bibliography{ref}

\begin{thebibliography}{10}
\providecommand{\url}[1]{#1}
\csname url@samestyle\endcsname
\providecommand{\newblock}{\relax}
\providecommand{\bibinfo}[2]{#2}
\providecommand{\BIBentrySTDinterwordspacing}{\spaceskip=0pt\relax}
\providecommand{\BIBentryALTinterwordstretchfactor}{4}
\providecommand{\BIBentryALTinterwordspacing}{\spaceskip=\fontdimen2\font plus
\BIBentryALTinterwordstretchfactor\fontdimen3\font minus \fontdimen4\font\relax}
\providecommand{\BIBforeignlanguage}[2]{{%
\expandafter\ifx\csname l@#1\endcsname\relax
\typeout{** WARNING: IEEEtran.bst: No hyphenation pattern has been}%
\typeout{** loaded for the language `#1'. Using the pattern for}%
\typeout{** the default language instead.}%
\else
\language=\csname l@#1\endcsname
\fi
#2}}
\providecommand{\BIBdecl}{\relax}
\BIBdecl

\bibitem{beth2024aipower}
``Ai power consumption: Rapidly becoming mission-critical,'' 2024, https://www.forbes.com/sites/bethkindig/2024/06/20/ai-power-consumption-rapidly-becoming-mission-critical/.

\bibitem{kann2024hear}
``Hearings start today on georgia power’s plan to burn more fossil fuels,'' 2024, https://www.ajc.com/news/hearings-start-today-on-georgia-powers-plan-to-burn-more-fossil-fuels/OTSUUETHNBCJ7JGSFIWEZ5MBQQ/.

\bibitem{baruch2019little}
G.~Baruch, M.~Baruch, and Y.~Goldberg, ``A little is enough: Circumventing defenses for distributed learning,'' \emph{Advances in Neural Information Processing Systems}, vol.~32, 2019.

\bibitem{yin2018byzantine}
D.~Yin, Y.~Chen, R.~Kannan, and P.~Bartlett, ``Byzantine-robust distributed learning: Towards optimal statistical rates,'' in \emph{International conference on machine learning}.\hskip 1em plus 0.5em minus 0.4em\relax Pmlr, 2018, pp. 5650--5659.

\bibitem{shejwalkar2021manipulating}
V.~Shejwalkar and A.~Houmansadr, ``Manipulating the byzantine: Optimizing model poisoning attacks and defenses for federated learning,'' in \emph{NDSS}, 2021.

\bibitem{neftci2019surrogate}
E.~O. Neftci, H.~Mostafa, and F.~Zenke, ``Surrogate gradient learning in spiking neural networks: Bringing the power of gradient-based optimization to spiking neural networks,'' \emph{IEEE Signal Processing Magazine}, vol.~36, no.~6, pp. 51--63, 2019.

\bibitem{venkatesha_federated_2021}
Y.~Venkatesha, Y.~Kim, L.~Tassiulas, and P.~Panda, ``Federated learning with spiking neural networks,'' \emph{IEEE Transactions on Signal Processing}, vol.~69, pp. 6183--6194, 2021.

\bibitem{aouedi_hfedsnn_2023}
\BIBentryALTinterwordspacing
O.~Aouedi, K.~Piamrat, and M.~Sûdholt, ``{HFedSNN}: Efficient hierarchical federated learning using spiking neural networks,'' in \emph{Proceedings of the Int'l {ACM} Symposium on Mobility Management and Wireless Access}.\hskip 1em plus 0.5em minus 0.4em\relax {ACM}, 2023, pp. 53--60. [Online]. Available: \url{https://dl.acm.org/doi/10.1145/3616390.3618288}
\BIBentrySTDinterwordspacing

\bibitem{yu2024heterogeneous}
Y.~Yu, Y.~Yan, J.~Cai, and Y.~Jin, ``Heterogeneous federated learning with convolutional and spiking neural networks,'' \emph{arXiv preprint arXiv:2406.09680}, 2024.

\bibitem{xie_efficient_2022}
\BIBentryALTinterwordspacing
K.~Xie, Z.~Zhang, B.~Li, J.~Kang, D.~Niyato, S.~Xie, and Y.~Wu, ``Efficient federated learning with spike neural networks for traffic sign recognition,'' \emph{IEEE Transactions on Vehicular Technology}, vol.~71, no.~9, pp. 9980--9992, 2022. [Online]. Available: \url{https://ieeexplore.ieee.org/document/9784851/}
\BIBentrySTDinterwordspacing

\bibitem{shah2021model}
S.~M. Shah and V.~K. Lau, ``Model compression for communication efficient federated learning,'' \emph{IEEE Transactions on Neural Networks and Learning Systems}, vol.~34, no.~9, pp. 5937--5951, 2021.

\bibitem{haddadpour2021federated}
F.~Haddadpour, M.~M. Kamani, A.~Mokhtari, and M.~Mahdavi, ``Federated learning with compression: Unified analysis and sharp guarantees,'' in \emph{International Conference on Artificial Intelligence and Statistics}.\hskip 1em plus 0.5em minus 0.4em\relax PMLR, 2021, pp. 2350--2358.

\bibitem{chaki_communication_2023}
\BIBentryALTinterwordspacing
S.~Chaki, D.~Weinberg, and A.~Özcelikkale, ``Communication trade-offs in federated learning of spiking neural networks.'' [Online]. Available: \url{http://arxiv.org/abs/2303.00928}
\BIBentrySTDinterwordspacing

\bibitem{liu_federal_2022}
\BIBentryALTinterwordspacing
Z.~Liu, Q.~Zhan, X.~Xie, B.~Wang, and G.~Liu, ``Federal {SNN} distillation: A low-communication-cost federated learning framework for spiking neural networks,'' in \emph{Journal of Physics: Conference Series}, vol. 2216, no.~1, 2022, p. 012078. [Online]. Available: \url{https://iopscience.iop.org/article/10.1088/1742-6596/2216/1/012078}
\BIBentrySTDinterwordspacing

\bibitem{xie2024federated}
X.~Xie, J.~Feng, G.~Liu, Q.~Zhan, Z.~Liu, and M.~Zhang, ``Federated learning for spiking neural networks by hint-layer knowledge distillation,'' \emph{Applied Soft Computing}, vol. 163, p. 111901, 2024.

\bibitem{fu_spikewhisper_2024}
H.~Fu, G.~Li, J.~Wu, J.~Li, X.~Lin, K.~Zhou, and Y.~Liu, ``Spikewhisper: Temporal spike backdoor attacks on federated neuromorphic learning over low-power devices.''

\bibitem{walter2024mitigating}
K.~Walter, M.~Mohammady, S.~Nepal, and S.~S. Kanhere, ``Mitigating distributed backdoor attack in federated learning through mode connectivity,'' in \emph{Proceedings of the 19th ACM Asia Conference on Computer and Communications Security}, 2024, pp. 1287--1298.

\bibitem{eshraghian2023training}
J.~K. Eshraghian, M.~Ward, E.~O. Neftci, X.~Wang, G.~Lenz, G.~Dwivedi, M.~Bennamoun, D.~S. Jeong, and W.~D. Lu, ``Training spiking neural networks using lessons from deep learning,'' \emph{Proceedings of the IEEE}, 2023.

\bibitem{lee2016training}
J.~H. Lee, T.~Delbruck, and M.~Pfeiffer, ``Training deep spiking neural networks using backpropagation,'' \emph{Frontiers in neuroscience}, vol.~10, p. 508, 2016.

\bibitem{bohte2011error}
S.~M. Bohte, ``Error-backpropagation in networks of fractionally predictive spiking neurons,'' in \emph{International conference on artificial neural networks}.\hskip 1em plus 0.5em minus 0.4em\relax Springer, 2011, pp. 60--68.

\bibitem{mcmahan2017communication}
B.~McMahan, E.~Moore, D.~Ramage, S.~Hampson, and B.~A. y~Arcas, ``Communication-efficient learning of deep networks from decentralized data,'' in \emph{Artificial intelligence and statistics}.\hskip 1em plus 0.5em minus 0.4em\relax PMLR, 2017, pp. 1273--1282.

\bibitem{li2021byzantine}
S.~Li, E.~Ngai, and T.~Voigt, ``Byzantine-robust aggregation in federated learning empowered industrial iot,'' \emph{IEEE Transactions on Industrial Informatics}, vol.~19, no.~2, pp. 1165--1175, 2021.

\bibitem{xie2020fall}
C.~Xie, O.~Koyejo, and I.~Gupta, ``Fall of empires: Breaking byzantine-tolerant sgd by inner product manipulation,'' in \emph{Uncertainty in Artificial Intelligence}.\hskip 1em plus 0.5em minus 0.4em\relax PMLR, 2020, pp. 261--270.

\bibitem{shi2019understanding}
S.~Shi, X.~Chu, K.~C. Cheung, and S.~See, ``Understanding top-k sparsification in distributed deep learning,'' \emph{arXiv preprint arXiv:1911.08772}, 2019.

\bibitem{nguyen2024robustness}
M.~V. Nguyen, L.~Zhao, B.~Deng, W.~Severa, H.~Xu, and S.~Wu, ``The robustness of spiking neural networks in communication and its application towards network efficiency in federated learning,'' \emph{arXiv preprint arXiv:2409.12769}, 2024.

\bibitem{li2024blades}
S.~Li, E.~C.-H. Ngai, F.~Ye, L.~Ju, T.~Zhang, and T.~Voigt, ``Blades: A unified benchmark suite for byzantine attacks and defenses in federated learning,'' in \emph{2024 IEEE/ACM Ninth International Conference on Internet-of-Things Design and Implementation (IoTDI)}.\hskip 1em plus 0.5em minus 0.4em\relax IEEE, 2024, pp. 158--169.

\end{thebibliography}

\end{document}